\documentclass[twocolumn,twoside,slac_two]{revtex4}
\usepackage{graphicx}
\usepackage{fancyhdr}
\begin{document}

\title{Measurement of the Charge Ratio of Atmospheric Muons with the CMS Detector}

%

\author{Gavin Hesketh, on behalf of the CMS Collaboration}
\affiliation{CERN, PH Dept, Meyrin, Switzerland}

\begin{abstract}
This paper describes a new measurement of the flux ratio of positive and negative muons from cosmic-ray interactions in the atmosphere, using data collected by the CMS detector at ground level and in the underground experimental cavern. The excellent performance of the CMS detector allowed detection of muons in the momentum range from 3 GeV to 1 TeV. For muon momenta below 100 GeV the flux ratio is measured to be a constant $1.2766 \pm 0.0032(stat) \pm 0.0032(syst)$, the most precise measurement to date. At higher momenta an increase in the charge asymmetry is observed, in agreement with models of muon production in cosmic-ray showers and compatible with previous measurements by deep-underground experiments.
\end{abstract}

\maketitle

\thispagestyle{fancy}

\section{Introduction}
This paper describes a measurement of the ratio the flux of positive to negative muons produced by cosmic ray interactions in the atmosphere, using the CMS detector~\cite{PLB, MUO-10-001}.
Cosmic ray particles (predominantly protons) interact with the upper atmosphere, initiating a hadronic shower. 
Kaons and pions produced in these showers may decay to muons, which can then reach the Earth's surface.
These muons are therefore an easily detectable signal of a cosmic ray interaction, and their properties provide information on the initial cosmic ray particle and development of the hadronic shower in the atmosphere.

The momentum distribution of muons has been parameterised~\cite{gaisser} based on the kinematics of primary cosmic-ray particles and of decays of secondary particles,
and this parametrisation can be used as the base to extract the muon charge ratio, $R$, as described in~\cite{arXiv09063726}:
\begin{equation}\label{cr_equation}
R = \frac{\frac{f_\pi}{1+1.1 E_\mu \cos\theta_z/115 \mathrm{GeV}}
        + \frac{\eta\cdot f_K}{1+1.1 E_\mu \cos\theta_z/850 \mathrm{GeV}} }
        {\frac{1-f_\pi}{1+1.1E_\mu \cos\theta_z/115 \mathrm{GeV}}
        + \frac{\eta\cdot(1- f_K)}{1+1.1 E_\mu \cos\theta_z/850 \mathrm{GeV}} }
\label{eq:PiKa}
\end{equation}

where the constants $f_\pi$ and $f_K$ are the fractions of all pions and kaons that are positive, and must be determined from data. The constant $\eta$ sets the relative pion and kaon fractions in cosmic-ray showers and the accepted value of 0.054~\cite{gaisser} is taken.
There is also a dependence on the vertical component of the muon energy, $E_\mu cos\theta_z$, where $\theta_z$ is the zenith angle.
Finally, there are two energy scales: 115~GeV and 850~GeV, above which pions and kaons respectively have a greater probability to interact rather than decay in the atmosphere.
Between 115 and 850~GeV therefore, the importance of kaons as a source of muons is expected to rise. 
As the value of $f_K$ is larger than $f_\pi$ due to the preference of $K^+\Lambda$ production over $K^+K^-$ production, the muon charge ratio is also expected to rise in the energy range between approximately 100~GeV and 1~TeV. 
Previous measurements of the muon charge ratio~\cite{utah,baxendale,rastin,hebbeker,L3C,Adamson:2007ww,arXiv09063726,OPERA} have been sensitive to energies below or above this expected rise, and this paper describes the first measurement to bridge this gap, while also being the most precise to date in the range 5-100~GeV.
Results are presented in terms of both momentum and the vertical component of the momentum, $p.cos\theta_z$.

\begin{figure}
\includegraphics[width=0.49\textwidth]{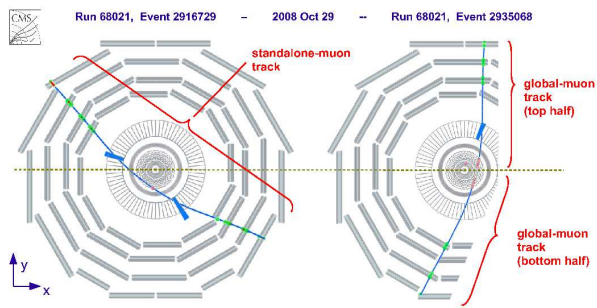}
\caption{Two cosmic ray muons reconstructed by CMS. Left: using only the muon detectors; right: using the muon and tracking detectors.
\label{fig:muon_reco}}
\end{figure}

The CMS detector~\cite{cms}, one of the experiments at the CERN Large Hadron Collider~\cite{lhc}, is designed to study proton-proton collisions at energies up to 14 TeV, and search for new physics at this energy frontier.
The detector is described in detail elsewhere~\cite{phys_tdr}, with the main feature being a superconducting solenoid magnet with a diameter of 6~m, providing a field of 3.8~T. Inside the magnet are a central tracking detector, surrounding the luminous region, and a crystal electromagnetic calorimeter and brass-scintillator hadronic calorimeter.
Muons are measured in gas-ionisation detectors embedded in the steel return yoke for the magnet.  
The ability to accurately measure TeV-energy muons is central to the physics goals of the LHC programme, with the added benefit that CMS also makes an excellent high-energy cosmic muon detector.
Figure~\ref{fig:muon_reco} shows two typical cosmic muon events reconstructed by CMS: using only the muon detectors, and a combination of muon and tracking detectors.

The data used in this analysis were collected in two phases of the commissioning of CMS: the 2006 ``Magnet Test and Cosmic Challenge'' (MTCC) runs~\cite{mtcc}, and the 2008 ``Cosmic Runs at 4 Tesla'' (CRAFT08)~\cite{CRAFT08}.
These two datasets were analysed separately due to the different detector conditions, and the results then combined.
Much of the work in designing the analyses focused on ensuring the cancellation of many charge-dependent systematic and acceptance effects, simplifying the analysis and reducing the overall uncertainties. 
This was complicated by the fact that, unlike some other experiments, CMS does not change the polarity of the magnet.

\section{MTCC Analysis}

\begin{figure}
\centering
   \includegraphics[width=0.49\textwidth]{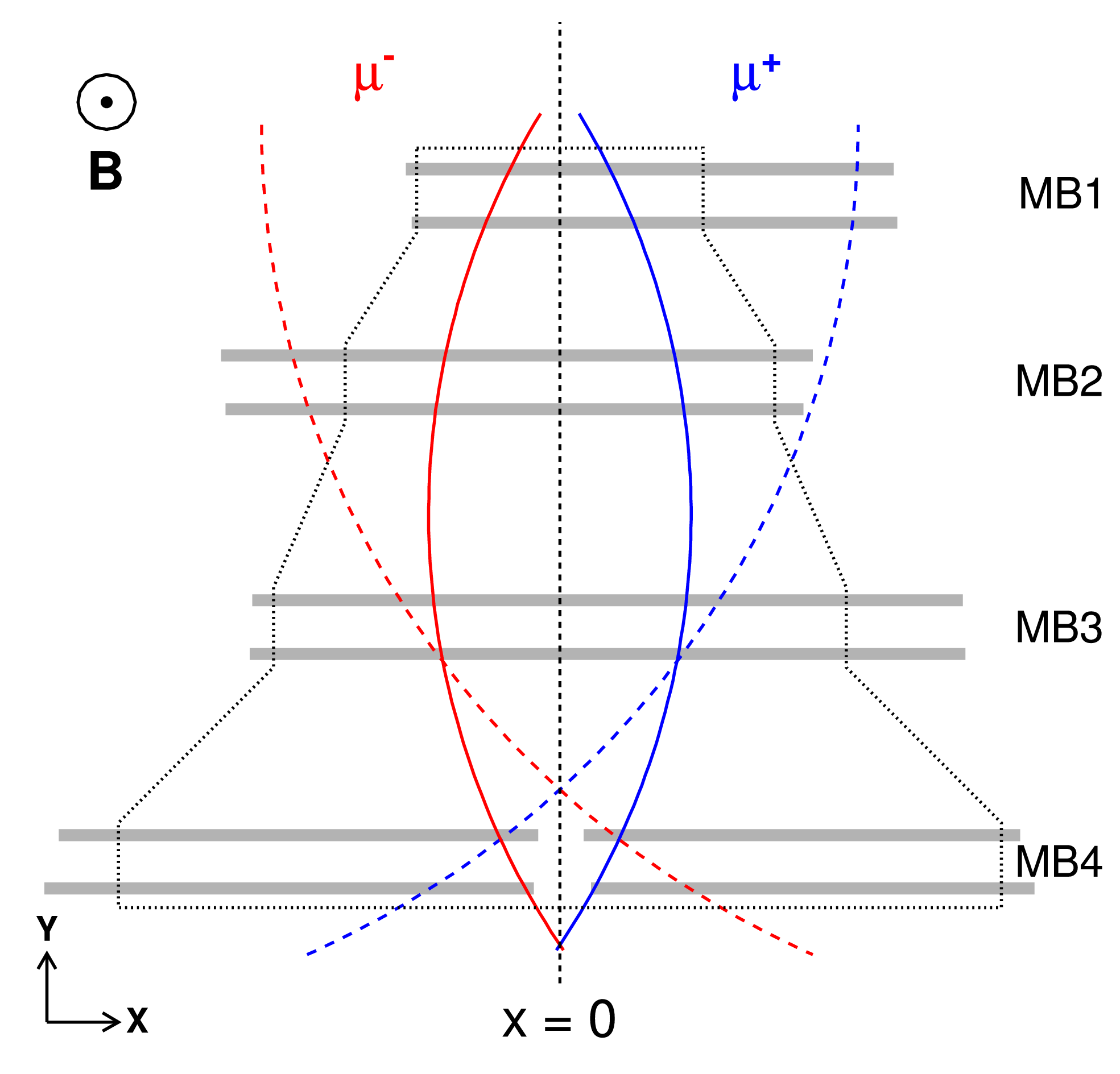}
\caption{Fiducial region for the MTCC analysis, showing the charge-symmetric acceptance and trigger for muons. The solid red and blue lines show muons that would be accepted, the dashed lines rejected.
\label{fig:mtcc}}
\end{figure}

During the MTCC runs, a part of the CMS detector was installed at ground level in the assembly hall above the main experimental cavern, at 509~m above sea level, $46^o18.57'$ north, $6^o4.62'$ east.
While two sections of the muon system were available, for this analysis a fiducial volume was defined using only one of these sections, to ensuring a symmetric acceptance for positive and negative muons, as shown in Fig.~\ref{fig:mtcc}.
A total of 330,000 muons were selected for the analysis, based on a trigger using the MB2 and MB4 layers of the detector (see Fig.~\ref{fig:mtcc}).
As the MTCC analysis was at the Earth's surface, it was possible to detect muons with momenta down to 3~GeV.
Reconstructed muons were corrected for energy lost traversing the detector, and the charge ratio as a function of $p$ and $p.cos\theta_z$ calculated.
Finally, a further correction for charge mis-identification is derived from simulation (using the CORSIKA Monte Carlo~\cite{corsika} and GEANT~\cite{geant} detector simulation) to obtain the final distribution.

\section{CRAFT08 Analyses}
\begin{figure}
\centering
   \includegraphics[width=0.49\textwidth]{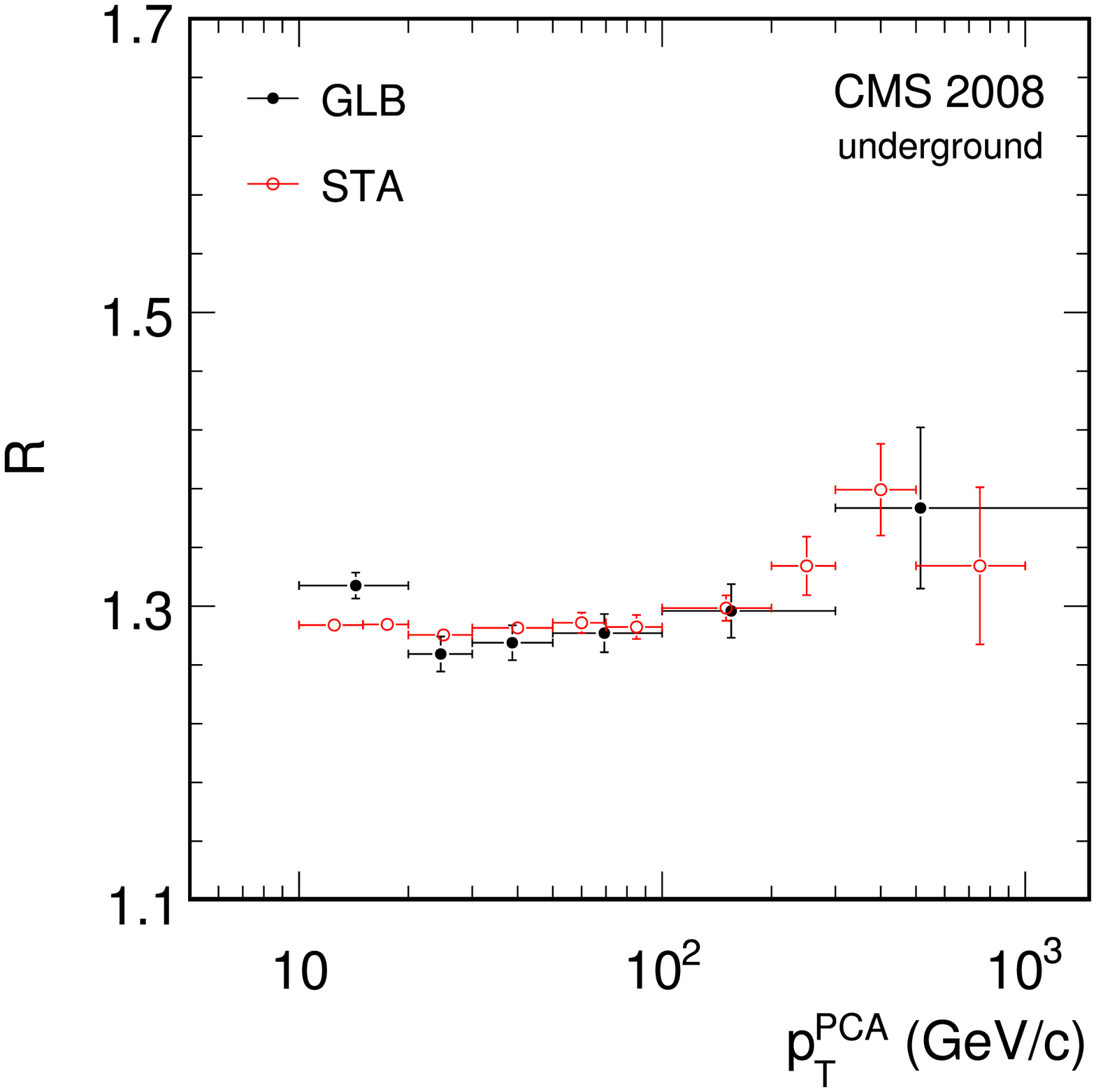}
\caption{The reconstructed charge ratios from the two CRAFT08 analyses, as a function of transverse momentum at the centre of the detector (p$_T^{PCA}$)} 
\label{fig:raw_ratios}
\end{figure}

For the CRAFT08 runs, the full CMS detector was installed in the experimental cavern, 89 m below the surface.
To identify cosmic ray muons traversing the detector (rather than muons produced in proton-proton collisions at the centre of the detector), special triggers and reconstruction algorithms were developed~\cite{CFT-09-014, L1triggerCraft}.
Two analyses were based on the data collected in these runs: one using ``global'' muons, and one using ``stand-alone'' muons. 
Global muons require the muons pass through the CMS muon system and central tracking system, which greatly improves the momentum resolution while greatly reducing the acceptance, relative stand-alone muons which use the muon system only.
A total of 300,000 events were selected for the global analysis, and 3~million for the stand-alone analysis.
In both cases, selections on the number of hits and the fit quality of the muons are used to minimise charge confusion, and events with more than one reconstructed muon rejected.
Figure~\ref{fig:raw_ratios} shows the charge ratio reconstructed in the detector for both analyses.

\begin{figure}
\includegraphics[height=0.49\textwidth,angle=90]{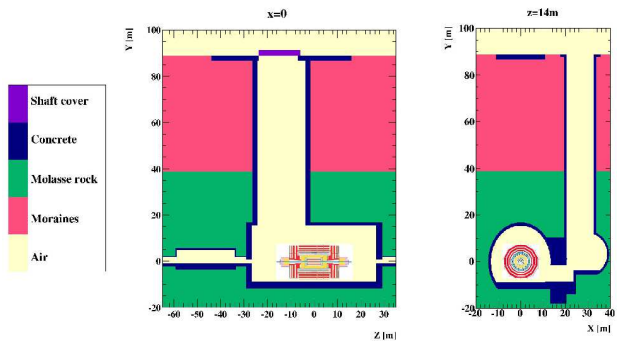}
\caption{Layout of the CMS experimental cavern, showing two of the three access shafts.
\label{fig:cms_cavern}}
\end{figure}

Both analyses then follow a similar method to extract the final result: the reconstructed muon tracks were individually propagated to the Earth's surface, correcting for energy loss both in the detector and in the rock above the cavern (the average energy loss being around 50~GeV).
However, leading down to the cavern are three access shafts, as shown in Fig~\ref{fig:cms_cavern}, and even very low energy muons can reach the detector by traveling down these shafts.
As two of the shafts are off the central axis of the detector, they break the acceptance symmetry for positive and negative muons: coming from one side of the detector, the solenoid magnet will bend positive muons into the detector acceptance, and negative muons out.
To overcome this, these shafts were excluded from the analysis, along with symmetric regions on the other side of the detector to preserve symmetry.

The distributions of $p$ and $p.cos\theta_z$ at the surface are then formed, and corrected for detector resolution and charge confusion effects.
The detector resolution for global muons was determined in data, using ``split tracks''. 
Muon tracks crossing the whole detector were split (see Fig.~\ref{fig:muon_reco}), and the resolution determined from the difference in the measurements of the incoming and outgoing legs.
For stand-alone muons, a GEANT based detector simulation was used to determine the resolution, which was cross-checked by comparing the stand-alone to central tracker resolution in the global muon data sample.

\section{Systematics}
For the MTCC analysis, the uncertainty on the detector alignment is the main limiting factor, at around 2\%, increasing with muon momentum to around 10\%. Much smaller uncertainties due to charge mis-identification and knowledge of the magnetic field also contribute.
For the CRAFT08 analyses, systematics were assessed due to the uncertainty on the material above the CMS cavern and any charge-dependent effect on the muon rate, the impact of the event selection and trigger, and knowledge of the magnetic field.
However, the largest single source of uncertainty was again the understanding of the detector alignment, which rises to 3-4\% at high momentum. 

The detector was aligned using cosmic muons during the MTCC and CRAFT08 runs~\cite{CFT-09-003}, however as well as uncertainties on this alignment, there remains the possibility of ``weak modes'': deformations of the detector which do not change the fitted track $\chi^2$, but do affect the measured track momenta.
As these weak modes are $\chi^2$ invariant, the standard alignment procedure is to some level blind to them, and the main concern for this analysis is the class of weak modes which induce a (signed) offset in the measured curvature, therefore shifting the measured momentum for positive and negatively charged muons in opposite directions.
To constrain such weak modes, the ``end-point test'' was developed.
The number of measured muons should fall to zero as the measured momentum goes to infinity, or as $q/p_T$ goes to zero. 
Fitting the measured $q/p_T$ distribution can test for any shift of this minimum away from zero, and set limits on the size of any weak modes. In the CRAFT08 analysis, an offset of $0.043\pm0.022~$TeV$^{-1}$ was found, the dominant uncertainty at high momentum.
It is worth noting that such weak modes will be further constrained by LHC collision data, as more alignment techniques become possible.

\section{Combination}

The fully corrected MTCC, global and stand-alone analyses were combined using a standard procedure~\cite{blue}.
Previous measurements have shown that the charge ratio is a constant in the range 5-100~GeV, so in a first combination all data points below $p.cos\theta_z=100$~GeV are combined into a single measurement.
This involves 6 points from the MTCC and 3 points from each of the global and stand-alone analyses and a full treatment of correlations between data points and analyses, yielding:
\begin{equation}
R = 1.2766 \pm 0.0032(stat) \pm 0.0032(syst)
\end{equation}
This is the most accurate measurement to date of the charge ratio in the range 5-100~GeV.

\begin{figure}
\includegraphics[height=0.49\textwidth]{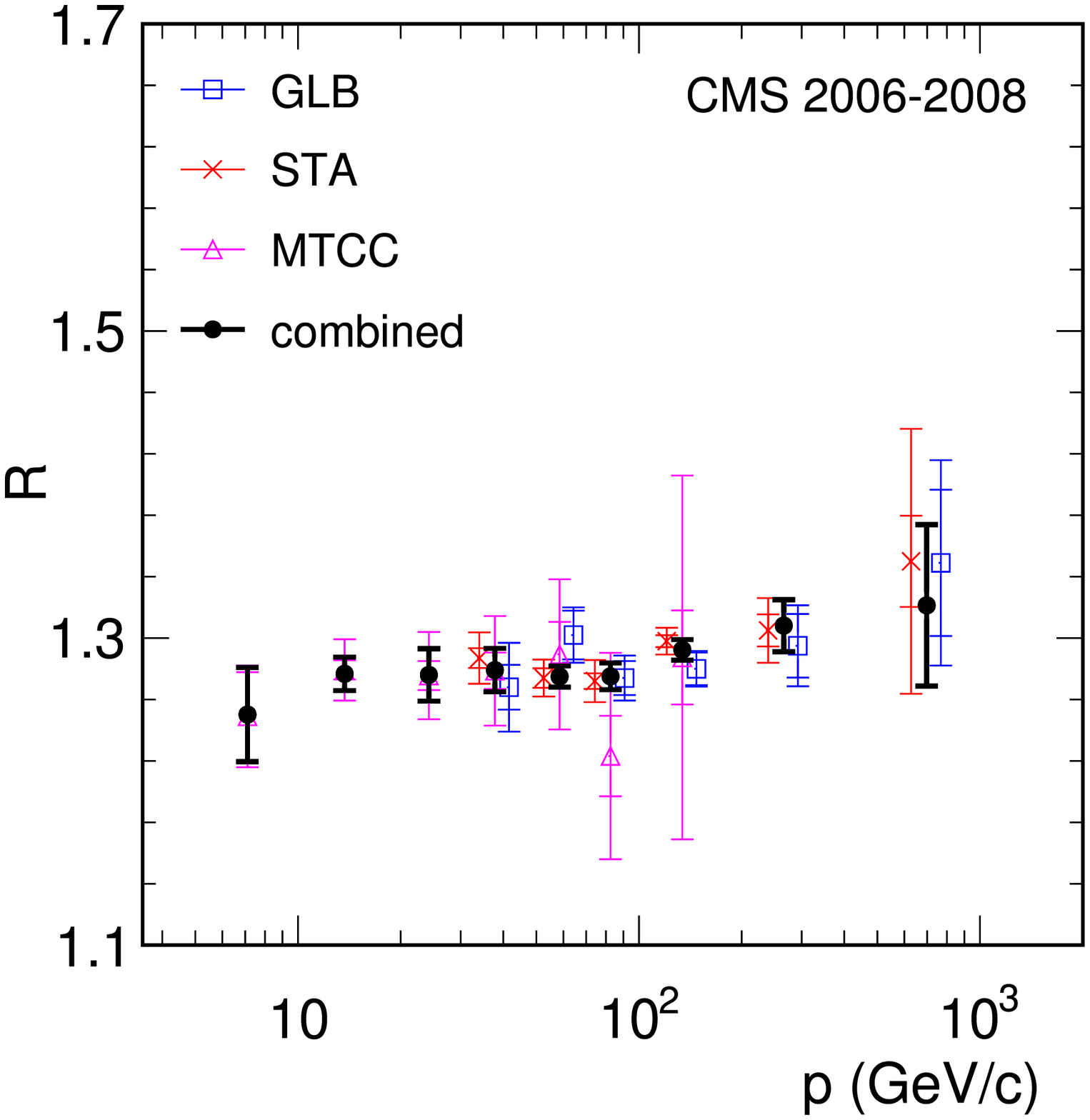}
\includegraphics[height=0.49\textwidth]{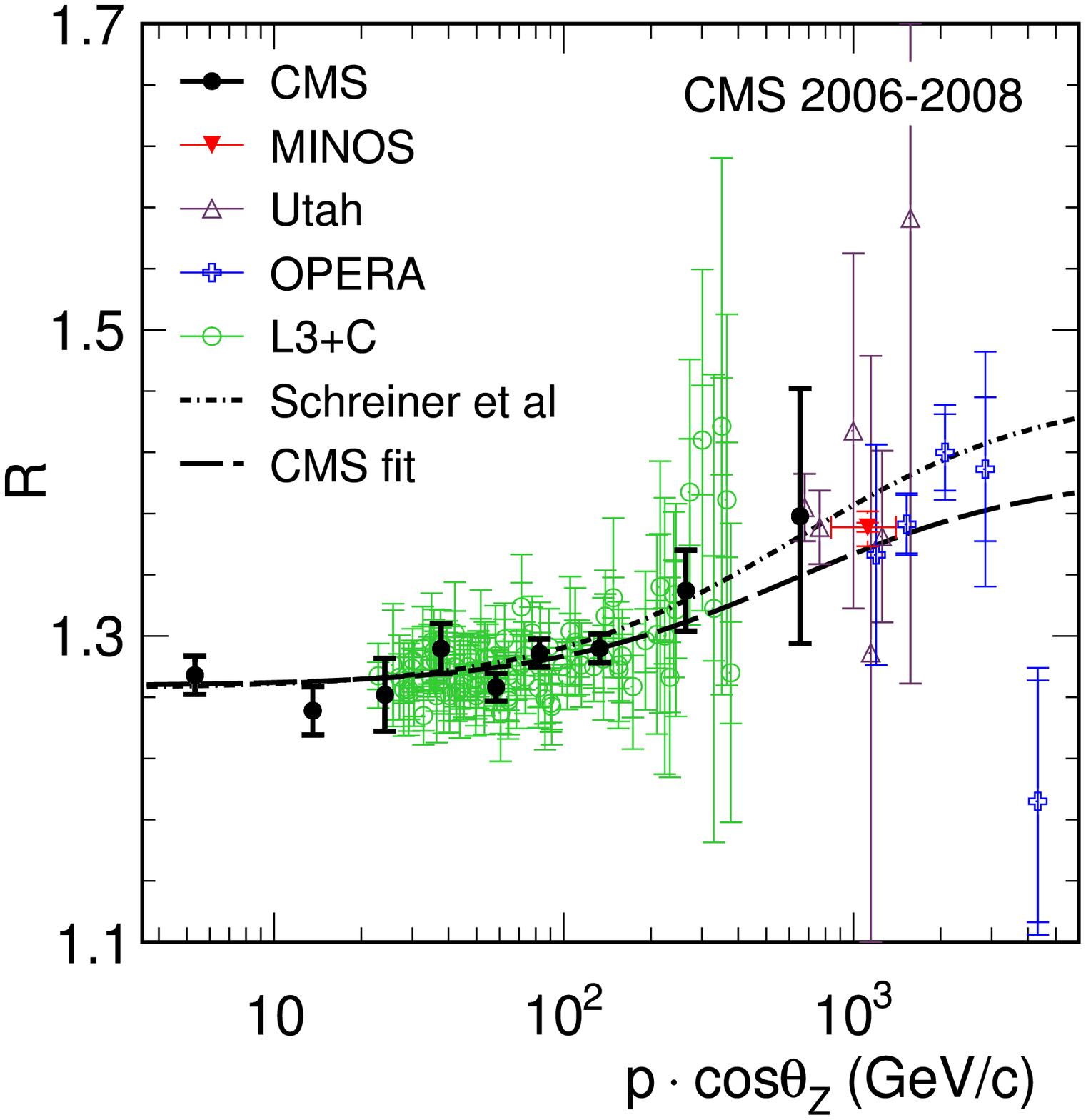}
\caption{Top: the three measurements of the charge ratio, R, and combination as a function of momentum, with the global and stand-alone analysis points offset for clarity.  Bottom: the CMS combination compared to previous measurements as a function of $p.cos\theta_z$.
\label{fig:combination}}
\end{figure}

Then, the data points from different analyses in each bin were combined, again keeping track of all correlations. 
The resulting combination is shown in Fig.~\ref{fig:combination}, and the charge ratio as a function of $p.cos\theta_z$ is also compared to previous measurements.
The CMS results agree well with the L3+C data~\cite{L3C} below 400~GeV, and with the Utah~\cite{utah}, MINOS~\cite{Adamson:2007ww} and OPERA~\cite{OPERA} measurements above 400~GeV.
The CMS data are also consistent with a rise in the charge ratio above 100~GeV, and fitting with Eq.~\ref{eq:PiKa} yields $f_\pi= 0.553 \pm 0.005$ and $f_K=0.66 \pm 0.06$, with a fit $\chi^2/d.o.f.= 7.8/7$.
It is worth noting that while the L3+C data contribute many points in the range 20-400~GeV, these points are in fact limited by a correlated systematic uncertainty. 
As a result, the CMS measurement is also the tightest constraint to date on $f_\pi$, which dominates the value of Eq.~\ref{eq:PiKa} below 100~GeV.

\bigskip

\section{Conclusions}
CMS has made a new measurement of the charge ratio of atmospheric muons. 
The measurement is the most precise to date in the range 5-100~GeV, and the first to span the interesting range 100-1000~GeV.
It is worth noting that the dominant systematic uncertainties, arising from the knowledge of detector alignment, will be reduced by LHC proton-proton collision data, making future cosmic ray muon measurements from CMS an interesting prospect.


\bibliography{gavin_hesketh_CMS_CR}

\end{document}